\documentclass[12pt,preprint]{aastex}

\usepackage{color}
    \definecolor{Blue}{rgb}{0.0,0.0,1.0}
    \definecolor{Green}{rgb}{0.0,1.0,0.0}
\usepackage[T1]{fontenc}

\slugcomment{submitted to ApJLett}

\shorttitle{Oscillations in outbursts}
\shortauthors{Li Xue et al.}

\begin{document}

\title{High frequency oscillations in outbursts of Kerr-metric slim disks}

\author{Li Xue\altaffilmark{1,2}}
\affil{Department of Astronomy and Institute of Theoretical Physics
and Astrophysics, Xiamen University, Xiamen, Fujian 361005, China}
\email{lixue@xmu.edu.cn}

\author{W{\l}odek Klu{\'z}niak\altaffilmark{2}}
\affil{Copernicus Astronomical Center, ul. Bartycka 18,
  PL-00-716 Warszawa, Poland}
\email{wlodek@camk.edu.pl}

\author{Aleksander S\k{a}dowski}
\affil{Harvard-Smithsonian Center for Astrophysics, 60 Garden St., Cambridge,
  MA 02134, USA}
\email{asadowski@cfa.harvard.edu}

\author{Ju-Fu Lu}
\affil{Department of Astronomy and Institute of Theoretical Physics
and Astrophysics, Xiamen University, Xiamen, Fujian 361005, China}
\email{lujf@xmu.edu.cn}

\and

\author{Marek Abramowicz\altaffilmark{1,2}}
\affil{Physics Department, Gothenburg University, SE-412-96
  G{\"o}teborg, Sweden}
\email{marek.abramowicz@physics.gu.se}

\altaffiltext{1}{Copernicus Astronomical Center, ul. Bartycka 18,
  PL-00-716 Warszawa, Poland}
\altaffiltext{2}{Institute of Physics, Faculty of Philosophy and
 Science, Silesian University in Opava, Bezru{\v c}ovo n{\'a}m. 13
 CZ-746-01 Opava, Czech Republic}

\begin{abstract}
We numerically investigate the thermally unstable accretion disks around
black holes. We adopt an evolutionary viscous stress equation to
replace the standard alpha-prescription based on the results of two MHD
simulations. We find a kind of interesting oscillations on
some running models in limit-cycle outburst state. The oscillations arise
near the inner boundary and propagate radially outwards. We deem that they
are the trapped $p$-mode oscillations excited by sonic-point instability. We directly
integrate the local radiation cooling fluxes to construct the mimic bolometric
light-curve. We find a series of overtones beside the fundamental harmonic on
the power spectra of mimic light-curves. The frequency of the fundamental
harmonic is very close to the maximum epicyclic frequency of the disk and the
frequency ratio of the fundamental harmonic and overtones is a regular integer series.
We suggest that the code for ray-tracing calculation must
be time-dependent in virtual observation and point out the robustness of the
black hole spin measurement with high frequency QPOs.
\end{abstract}


\section{Introduction}
High-frequency quasi-periodic oscillations (HFQPOs) have been observed in some
black hole (BH) X-ray binaries, which only appear in "steep power law" state
in high luminosity ($L>0.1L_{\rm{Edd}}$) and are in range of 40 to
$450\rm{Hz}$. These frequencies are comparable to the orbital frequency of
innermost stable circular orbit (ISCO) of a stellar-mass BH. It is believed
that the mechanism behind them is closely related to the dynamics of inner
regions of BH accretion disks
\citep[see][for reviews]{Remillard06,Kato_book,Belloni12}.

The pioneering work of \cite{Kato78} studied the effect of viscosity on the
disk stability and found the sonic-point instability (its original name is
pulsational instability), which is the excitation
process of so called $p$-mode oscillations in transonic accretion flows
(this oscillation has the other name, inertial-acoustic oscillation, but
we only call it $p$-mode oscillation in this paper). The
mechanism of this instability is analogous to the $\epsilon$-mechanism in
stellar pulsations. If the viscous parameter $\alpha$ of $\alpha$-prescription
\citep{SS73} is larger than a critical value, the instability will arise from
the phase relation between the viscous heat generation and oscillation,
i.e. the viscous heat generation increase (decrease) in the compressed
(expanded) phase of disk oscillation \citep{Kato78,Kato88}.

A series of works \citep{Matsumoto88,Matsumoto89,CT95,MHL14} were dedicated to
study the role of viscosity on HFQPOs. They all found the oscillation, whose
frequency is close to the maximum of epicyclic frequency $\kappa_{\rm{max}}$.
Additionally, \cite{MHL14} also found many overtones, whose frequencies are
close to the integer multiples of $\kappa_{\rm{max}}$.

Though the famous $\alpha$-prescription has been used extensively since 1973,
it may be too simple to closely accord with actual accretion flows. The
magnetohydrodynamic (MHD) shearing box simulations of \cite{Hirose09} implied
that there is certain time-delay between the viscous stress and total pressure
instead of the instantaneity introduced by $\alpha$-prescription. The other
work of \cite{Penna13}, which is based on a few relativistic MHD global
simulations, pointed out that the parameter $\alpha$ is a function of radius
but not constant in the inner disk region. More than one decade ago,
\cite{YK96} analyzed effects of viscous time-delay on the $p$-mode oscillations
and they found that it has a little of inhibition on the growth of
oscillations but excitation still exists for high viscosity cases.

Our work presented in this letter is also dedicated to study the role of
viscosity on QPOs. The distinction of our model is the evolutionary viscous
stress equation (instead of the $\alpha$-prescription), which is constructed
to mimic effects of viscous time-delay and inconstant $\alpha$ basing on the
results of \cite{Hirose09} and \cite{Penna13}.

\section{Equations}
In this letter, we consider the axisymmetric relativistic accretion flows
around black holes. The Boyer-Lindquist coordinates $t$, $r$, $\theta$, $\phi$
are used to describe the space-time. The governing equations of accretion
flows are written as following,
\begin{eqnarray}
&&\frac{\partial\Sigma}{\partial t} = -\frac{r\Delta^{1/2}}{\gamma A^{1/2}}
\left[\Sigma\frac{\partial u^t}{\partial t}+\frac{1}{r}\frac{\partial}{\partial
r}\left(r\Sigma\frac{V}{\sqrt{1-V^2}}\frac{\Delta^{1/2}}{r}\right)\right],
\label{continuity}  \\
&&\frac{\partial V}{\partial t} = \frac{\sqrt{1-V^2}\Delta}{\gamma A^{1/2}}
\left[-\frac{V}{1-V^2}\frac{\partial V}{\partial r}+\frac{{\cal A}}{r}
-\frac{1-V^2}{\rho}\frac{\partial p}{\partial r}\right], \label{radial} \\
&&\frac{\partial\mathcal{L}}{\partial t} = -\frac{\Delta}{\gamma A^{1/2}}
\frac{V}{\sqrt{1-V^2}}\frac{\partial\mathcal{L}}{\partial r}
-\frac{\Delta^{1/2}}{\gamma A^{1/2}\Sigma}\frac{\partial}{\partial r}
\left(\frac{2\Delta}{r} S_{r\phi}\right),\label{angular}\\
&&\frac{\partial H}{\partial t} = -\frac{UH}{r} -\frac{1}{\gamma}
\frac{V}{\sqrt{1-V^2}}\frac{\partial H}{\partial r},\label{thickness}\\
&&\nonumber \frac{\partial U}{\partial t} = \frac{\Delta^{1/2}r}{\gamma^2
A^{1/2}H}{\cal R}-\frac{U}{\gamma^2}\left(\frac{V}{(1-V^2)^2}
\frac{\partial V}{\partial t}+\frac{{\cal L}r^2}{A}
\frac{\partial\cal L}{\partial t}\right)\\
&&~~~~~-\frac{U}{H}\frac{\partial H}{\partial t}, \label{vertical}\\
&&\nonumber \frac{\partial T}{\partial t} = \frac{1}{\Sigma}
\frac{r\Delta^{1/2}}{\gamma A^{1/2}}\left[\frac{F^+-F^-}{c_V}
+(\Gamma_3-1)T\Sigma\left(-\frac{\partial u^t}{\partial t}
-\frac{1}{r^2}\frac\partial{\partial r}(r^2u^r)\right)\right]\\
&&~~~~~-\frac{V\Delta}{\gamma\sqrt{1-V^2}A^{1/2}}
\frac{\partial T}{\partial r}.\label{energy}
\end{eqnarray}
Where $\Sigma$, $V$, $\mathcal{L}$, $H$, $U$, $T$, and $S_{r\phi}$ are the
surface density, radial velocity (measured in the corotating frame), angular
momentum per unit mass ($\mathcal{L}\equiv u_{\phi}$), half thickness of disk,
vertical velocity of the surface, local temperature of accreted gas, and
viscous stress (only $r\phi$-component is non-vanishing), respectively.
These equations are all derived from the conservation of the stress-energy
tensor, and almost the same as those in our previous paper \cite{PaperII}
(see that paper for the detailed derivations as well as definitions of
$\gamma$, $A$, $\Delta$, and etc).

As mentioned in previous section, we adopt an additional evolutionary stress
equation to describe the viscosity instead of the $\alpha$-prescription. This
equation can be written as
\begin{equation}
n\tau^*\frac{\partial S_{r\phi}}{\partial t}=S^*_{r \phi}-S_{r \phi},
\label{stress}
\end{equation}
where the factor $n\tau^*$ is the practical viscous time-delay, which is
scaled with the typical delay $\tau^*$ by parameter $n$; $S^*_{r\phi}$ is the
expected stress by turbulence. The definitions of $\tau^*$ and $S^*_{r\phi}$
can be written as
\begin{eqnarray}
&&\tau^*=-\left(\frac{\gamma^2 A\Omega}{r^4}
\frac{\partial\ln\Omega}{\partial\ln r}\right)^{-1}, \label{def-tau} \\
&&S^*_{r\phi}=-\frac{\nu\Sigma A^{3/2}\gamma^3}{2r^3\Delta^{1/2}}
\frac{\partial\Omega}{\partial r}, \label{def-Srphi*}
\end{eqnarray}
where
\begin{eqnarray}
&&\Omega=\frac{d\phi}{dt}, \\
&&\nu=\frac{2}{3}\alpha H \sqrt{\frac{p}{\rho}}, \label{nu}\\
&&\alpha=\alpha_0\left(\frac{1-2Mr^{-1}+a^2r^{-2}}{1-3Mr^{-1}+2aM^{1/2}r^{-3/2}}
\right)^6. \label{alpha-prof}
\end{eqnarray}
If $\alpha=\rm{const}$ and $n\rightarrow 0$, equation (\ref{stress}) will
reduce to $S_{r\phi}=S^*_{r\phi}$, which is the same as the
$\alpha$-prescription in \cite{PaperII}. It means that equation (\ref{stress})
contains the $\alpha$-prescription as a trivial case.
Equation (\ref{alpha-prof}) determines the dependence of $\alpha$ on radius
$r$, BH mass $M$ and spin $a$. The radial factor, including the exponent $6$,
was suggested by \cite{Penna13}. Under this profile, $\alpha$ is almost a constant
$\alpha_0$ in outer disk region with large $r$ and increases to higher value radially
inwards. We set $\alpha_0=0.1$ and fix the mass supplying rate $\dot{M}_{\rm{out}}=0.06\dot{M}_{\rm{Edd}}$
(see table \ref{tab1} for the definition of Eddington accretion rate, $\dot{M}_{\rm{Edd}}$) at the outer boundary
in this work. In practice, these settings are sufficient to make the disk thermally unstable and
we indeed observe the limit-cycle outburst from running code.

After defining $S_{r\phi}$, the viscous heating rate $F^+$ in equation
(\ref{energy}) is redefined as
\begin{equation}
F^+=-2\frac{\gamma A^{1/2}\Delta^{1/2}}{r^3}\frac{\partial\Omega}{\partial r}
\cdot S_{r\phi}, \label{vis_heating}
\end{equation}
which would be reduced to the one in \cite{PaperII} when $n\rightarrow 0$.

\section{Instabilities}
We update our previous code established in \cite{PaperII} and run it for
eleven numerical models, whose parameters are listed in table \ref{tab1}.
\citet{Lin11} and \citet{Ciesielski12} studied the impact of viscous
time-delay on the thermal instability and found that it is not remarkable for
small enough time-delay. Indeed, we observe the expected limit-cycle
outbursts on all models with the time delay parameters $n$ in the range $0$ to
$4$. Therefore, the time-delay implemented in our code does not affect the
limit-cycle outbursts.

Among these models, S8 and S30 are two typical ones. S8 is a disk around a
non-spinning black hole as well as S30 around a fast-spinning one. In figure
\ref{Fig:LightCurves}, we show the bolometric light curves for these two
models respectively. For each point on light-curves, the disk luminosity is
made by integrating the local radiation cooling fluxes, and the sampling time
corresponds to the frequency $10^4\rm{Hz}$, which is enough to reveal any
harmonic with frequency lower than $5\times 10^3\rm{Hz}$ in the power spectral
density (PSD). Due to the difficulties in hydrodynamical calculation, we only
obtain one outburst light curve for S8 and $0.7\rm{s}$-long luminosity
ascending light curve for S30. However, they are long enough for the
calculation of PSDs. In figure \ref{Fig:PSDs}, we show the relevant PSDs for
S8 and S30. The fundamental frequency (the lowest frequency of harmonics) is
$\sim74.9\rm{Hz}$ for S8 and $\sim285.6\rm{Hz}$ for S30, which are both close
to $71.3\rm{Hz}$ and $300\rm{Hz}$, the respective maximal epicyclic frequencies,
which are also the maximum predicted frequencies of axisymmetric oscillations in
the trapped $p$-mode theory
\citep[see][and a detailed relativistic analysis is ongoing by our colleague Ji\u{r}\'{i} Hor\'{a}k]{Kato01,Kato_book}.
The spectrum of axisymmetric ($m=0$), horizontal $p$-modes
was recently computed by \cite{Giussani14}, who show that
in addition to a discrete set of lower frequency modes which
are trapped in the inner disk, there are modes of frequency
very close to the maximal epicyclic frequency
in which the oscillation is transmitted to the outer disk.
In table \ref{tab1}, we also list the fundamental
frequencies observed from the other oscillating models, which are all close to the
theoretical values though they vary in a narrow frequency range.

\begin{figure}
\centering
\includegraphics[width=0.6\textwidth]{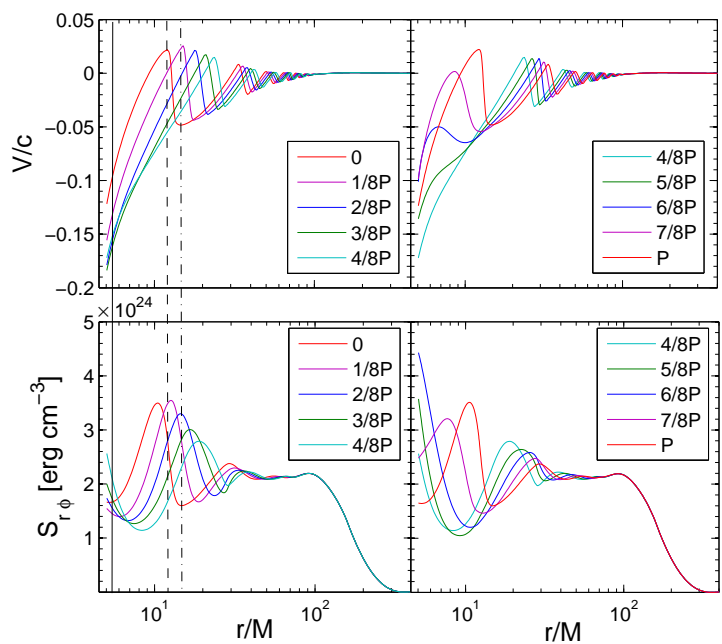}
\caption{One oscillation cycle of the radial velocity and viscous stress from model S8. The
different color lines denote a serial snapshotting times, which are all scaled
with the period of this cycle. The radius $r$ is scaled with BH mass $M$
because we take $G=c=1$. Thus the Schwarzschild radius is $2M$.}
\label{Fig:OneCycle}
\end{figure}

In figure \ref{Fig:OneCycle}, we show an oscillating cycle of model S8 for $V$
(upper two panels) and $S_{r\phi}$ (lower two panels). The oscillations arise
near the inner boundary and propagate outwards. In the figure, one can follow 
the motion of individual wavelets. The negative
$V$ denotes inflow and negative gradient of $V$ corresponds to
the compression (inflow speed of inner is slower than outer) as well as positive
gradient to the expansion (inflow speed of inner is faster than outer). Thus, any wavelet on
the $V$-profiles can be divided into the compressed wave-front and expanded
wave-rear regions. For example, the left-most wavelet on initial $V$-profile (red
curve in the upper-left panel) can be divided into the wave-front (between dashed
and dash-dotted lines) and wave-rear (between solid and dashed lines) regions.
The relevant variation of $S_{r\phi}$, which is proportional to the viscous heating, is showed in
the lower-left panel. $S_{r\phi}$ (as well as the viscous heating) monotonously increases
in compressed wave-front (see the red curve between dashed and dash-dotted lines) but it unceasingly
increases after the maximal compression (takes place at the dashed line) and then decreases
in the expanded wave-rear. This non-monotonicity of $S_{r\phi}$ in wave-rear
is due to the time-delay contained by equation \ref{stress}. We deem that this is the phase
relation between the viscous heat generation and disk oscillation, which implies the arise of
sonic-point instability for $p$-mode oscillation \citep{Kato78,Kato88}. Thus, we also
deem that the oscillations actually arise near the sonic-point, which is included in
our computational domain and very close to the inner boundary.

\section{Results and Discussions}

In table \ref{tab1}, there are the other nine models with different viscous
settings but with the same BH mass, BH spin and mass supply rate (fixed
accretion rate at the outer boundary) as S8. We only observe oscillations on
the models with non-vanishing delay and large enough constant $\alpha$
($\gtrsim 0.3$) or $\alpha$-profile. In fact, the effective value on the
$\alpha$-profile increases inwards from constant $0.1$ in outer disk region
to the maximum $0.45$ at the inner boundary. Thus, the impact of
$\alpha$-profile is similar to the large constant $\alpha$ while it just
becomes large enough near the origin of instability, sonic-point. The facts
of these oscillating models imply that the large $\alpha$ (at least near
sonic-point) is a necessary condition for the arise of sonic-point
instability as well as $p$-mode oscillation, which is consistent with the
analysis of \cite{Kato88} and \cite{YK96}.

On the other view, we note the effect of delay on the appearance of
oscillations. The oscillating models with $n=1$ (S7 and S8) have the
oscillations during the whole outburst. The other oscillating models with
$n\neq 1$ (S1, S2 and S17) lose the oscillations in different luminosity
stages. The models with $n=0$ (S3, S4, and S16) have no any oscillation though
they have large enough $\alpha$, while the disappearance of the oscillations
on models S5 and S15 is due to the small $\alpha$. These facts imply that
$\tau^*$ may be a favorable delay for oscillation excitation on the disk
around a $10M_{\odot}$ Schwarzschild BH under the large $\alpha$. The last
model in table \ref{tab1}, S30 is a special case for a fast-spinning Kerr BH,
which has another favorable delay $4\tau^*$. This may imply the dependence
between the viscous time-delay and BH spin.

Focusing on the luminosity of the oscillating models, we observe oscillations
only in the limit-cycle outburst state ($L\gtrsim0.2L_{\rm{Edd}}$) when the
inner disk region has switched to slim disk mode. On the contrary, there is no
any oscillation observed in the limit-cycle quiet state
($L\thicksim 0.01L_{\rm{Edd}}$). This is consistent with the HFQPO
observations, but cannot be compared with the sonic-point instability theory
which does not discriminate between accretion rates. Recently, the shearing
box simulation of \cite{Hirose14} implied that the effective $\alpha$ is
enhanced by the vertical convection during the outburst, which is similar to
the conception of \cite{MCT94}. Thus, larger $\alpha$ required by the $p$-mode
oscillation may be caused by the outburst, explaining why HFQPOs are observed
only in high luminosity state.

\begin{figure}
\centering
\includegraphics[width=0.6\textwidth]{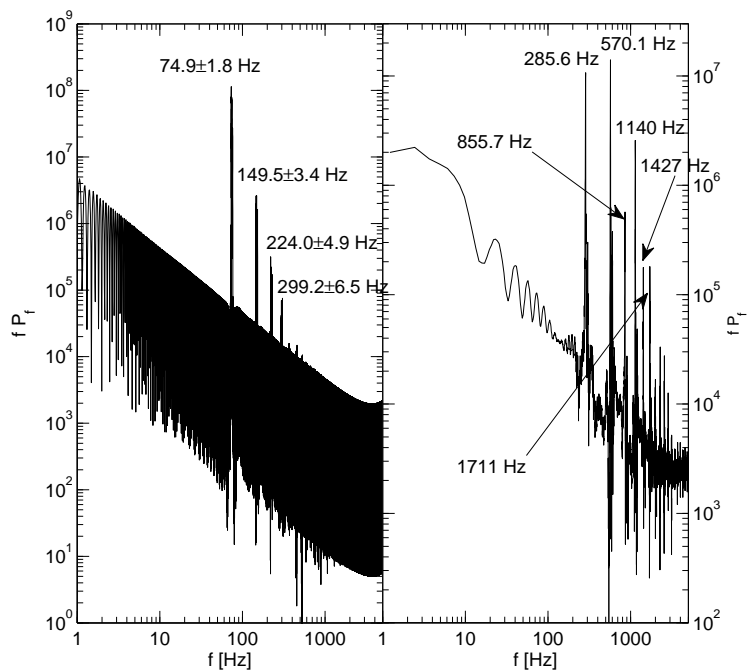}
\caption{PSDs of models S8 (left) and S30 (right).}
\label{Fig:PSDs}
\end{figure}

Beside the fundamental harmonic, there are many overtones in both of the two
spectra in figure \ref{Fig:PSDs}. The frequency ratio of fundamental
harmonic and its overtones is a regular integer series, which is also observed
by \cite{MHL14} in their 2D-simulations (in radial and azimuthal dimensions)
when the axisymmetric $p$-mode become dominant for large $\alpha$. However,
no overtones are observed by \cite{CT95}. Perhaps, this is due
to their adopting of the viscous prescription with
$S_{r\phi}\propto p_{\rm{gas}}$ instead of $S_{r\phi}\propto p_{\rm{total}}$
in \cite{MHL14} and our models. These interesting overtones may be potentially
useful for explaining the observational QPO pairs, which are always in a specific
integer frequency ratio.


Subsequently, the virtual observation from our numerical results will be
logical and interesting. However, it would require more careful treatment. For
example, the effective time-delay of arrival caused by the gravitational
bending on the trajectories of emitted photons and the large observational
view-angle, and effective blocking caused by the gravitational red-shift and
the other shields.

\begin{figure}
\centering
\includegraphics[width=0.6\textwidth]{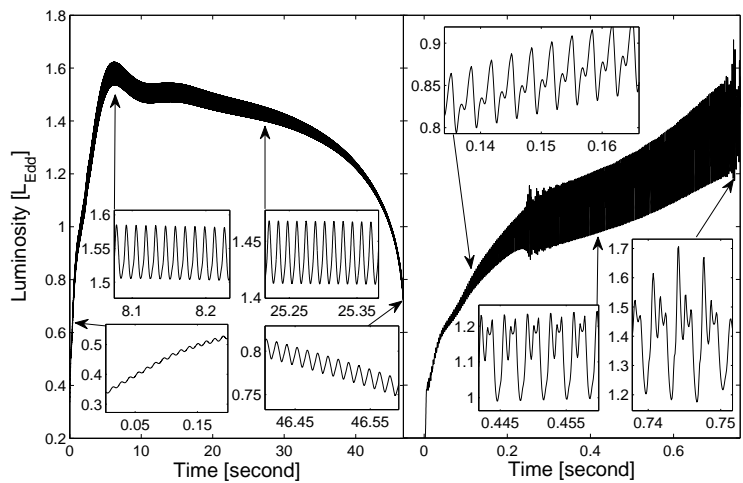}
\caption{Light-curves of models S8 (left) and S30 (right). There are a few
subplots to reveal detailed views of surrounding arrow points.}
\label{Fig:LightCurves}
\end{figure}

As a rough evaluation for the effective time-delay of arrival, one can
consider the observation from an almost edge-on disk, on which the photons
emitted from two locations apart from distant $\Delta r=203M$ at the same
observer's time ($t$) will arrive at the observer with the rough delay
$\Delta t=0.01\rm{s}$ comparable with the period of $71.3\rm{Hz}$ (our results
presented in figures \ref{Fig:LightCurves} and \ref{Fig:PSDs} can be roughly
regarded as from the face-on disks). The distant $\Delta r=203M$ is comparable
with the radius of outward wavelets propagating area, so it is possible to
change observer's final view on the oscillation power spectrum. This also
implies that the code for ray-tracing calculation on virtual observation must
be time-dependent.

\begin{figure}
\centering
\includegraphics[width=0.6\textwidth]{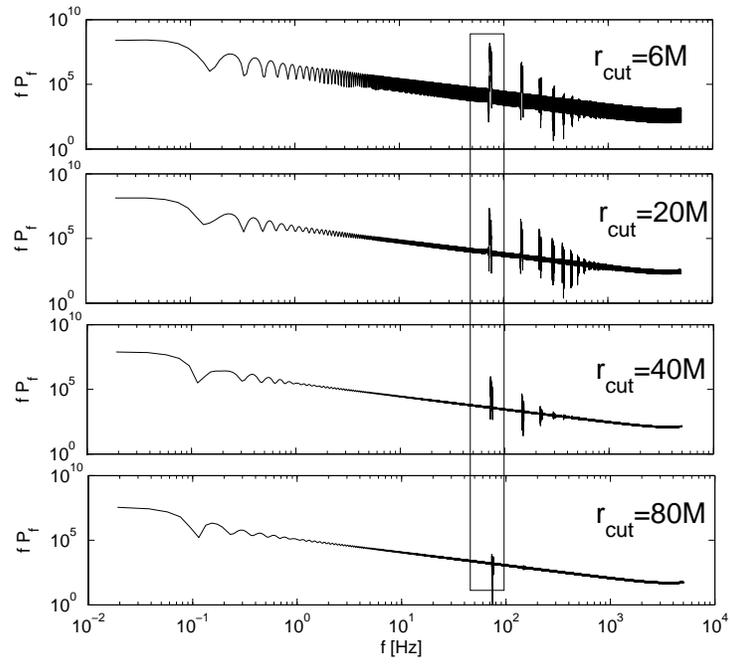}
\caption{PSDs of different blocking cases. From upper to lower, the radii of
cutting regions $r_{\rm{cut}}$ increase for the same model S8.}
\label{Fig:4PSDs}
\end{figure}

In order to roughly demonstrate the effective blocking, we calculate the
light-curves without the radiation contribution from different inner cutting
regions for the same model S8 and we show the relevant power spectra in
figure \ref{Fig:4PSDs}. It is remarkable that the fundamental harmonic (inside
the rectangle in all four panels) cannot be easily removed from PSDs because
of the outward propagation of the oscillation from sonic-point. It implies
that the measurement of BH spin with HFQPO will be very robust even in a case
when modulation of the innermost disk is not visible.

So far, we have found some features of our model fortunately coinciding with
the counterparts of HFQPO observations. All of these are only due to the
adopting of a special evolutionary stress equation in our model, which mimics
the viscous features induced from MHD simulations. However, our model still
lacks some abilities for capturing various complicated features associated
with the real accretion flows around BHs. In fact, it is almost  impossible
for seeing the oscillations during the whole outburst state (like those in
figure \ref{Fig:LightCurves}) in real accretion flows. It is because the
effective $\alpha$ and viscous time-delay, determining the appearance of
oscillations, is turbulent stochastic in real accretion flows though their
expected values can be determined by certain laws. Further more MHD
simulations on the mean behaviors of effective turbulent viscosity is
necessary for improving our understanding on accretion process around BHs
\cite[e.g.][]{Hirose09,Hirose14,Penna13}. It is also impossible for capturing
the resonance between the radial and vertical oscillations suggested by
\cite{Kluzniak04}, because our model is a vertically integrated model and
there is only radial dependence reserved. The further analytic works and MHD
simulations on the roles of radial and vertical oscillations are necessary for
explaining the observational HFQPO pairs though our model has the intrinsic
multi-frequency feature. In fact, the thermal instability required by our
model is the theoretical one whose deviations from the observations have been
found a decade ago \citep{GD04}. While the existence of thermal instability on
the BH accretion disks is still an open issue at present, we only can adopt
this theoretical thermal instability to produce the high luminosity outburst
required by the oscillations in our model. We believe the plausibility of
oscillations observed on our models though it is "dancing" on a poor-quality
stage.

\acknowledgements
This work was supported by the National Natural Science Foundation of China
under grants 11233006 and 11373002, Polish NCN grant UMO-2011/01/B/ST9/05439 
and 2013/08/A/ST9/00795, and Czech ASCRM100031242 CZ.1.07/2.3.00/20.0071 Synergy 
(Opava) project.


\clearpage

\begin{deluxetable}{cccccccc}
\tabletypesize{\scriptsize}
\tablecolumns{8}
\tablecaption{Model Sequences\label{tab1}}
\tablewidth{0pt}
\tablehead{&&&&&& \colhead{Fundamental} & \colhead{Max Epicyclic}\\
\colhead{ID}\vspace{-0.1cm} & \colhead{$n$}
 & \colhead{$\alpha$} & \colhead{$M/M_\odot$}
 & \colhead{$a/M$} & \colhead{$\dot{M}/\dot{M}_{\rm{Edd}}$\tablenotemark{1}}
 & \colhead{Frequency [Hz]} & \colhead{Frequency [Hz]}}
\startdata
S8 & 1    & Eq. (\ref{alpha-prof}) & 10   & 0     & 0.06 & 71.64-76.73\tablenotemark{2}  & 71.3\\
S15& 1    & 0.1                    & 10   & 0     & 0.06 & No Osci.     & 71.3 \\
S5 & 1    & 0.15                   & 10   & 0     & 0.06 & No Osci.     & 71.3\\
S7 & 1    & 0.3                    & 10   & 0     & 0.06 & 62.37-68.7   & 71.3\\
S4 & 0    & 0.3                    & 10   & 0     & 0.06 & No Osci.     & 71.3\\
S3 & 0    & 0.45                   & 10   & 0     & 0.06 & No Osci.     & 71.3\\
S16& 0    & Eq. (\ref{alpha-prof}) & 10   & 0     & 0.06 & No Osci.     & 71.3\\
S2 & 0.5  & Eq. (\ref{alpha-prof}) & 10   & 0     & 0.06 & 74.77-80.64 (AD)\tablenotemark{3} & 71.3\\
S17& 0.75 & Eq. (\ref{alpha-prof}) & 10   & 0     & 0.06 & 74.27-78.5 (AD)  & 71.3\\
S1 & 2    & Eq. (\ref{alpha-prof}) & 10   & 0     & 0.06 & 69.77-71 (P)   & 71.3\\
\\
S30& 4    & Eq. (\ref{alpha-prof}) & 7.02 & 0.947 & 0.06 & 285.64 \tablenotemark{4}   & 300\\
\enddata
\tablenotetext{1}{Eddington accretion rate
$\dot{M}_{\rm{Edd}}\equiv\frac{64\pi GM}{c\kappa_{\rm{es}}}=2.23\times 10^8\frac{M}{M_{\odot}}\rm{(g s^{-1})}$.}
\tablenotetext{2}{The variation range of observed fundamental frequencies is given.}
\tablenotetext{3}{The characters inside the parentheses denote the luminosity
stages in which the oscillations disappear. For example, (AD) denotes the
oscillations only appear in the luminosity \emph{plateau} stage (disappear in
\emph{ascending} and \emph{descending} stages).}
\tablenotetext{4}{We only have the data for the luminosity ascending stage.
Due to the lack of data, we cannot observe any remarkable variance on its
fundamental frequency.}
\end{deluxetable}

\clearpage


\end{document}